\documentclass[aps,pra,amsmath,amssymb,floatfix,twocolumn,amsmath,superscriptaddress,twocolumn,nofootinbib,tighten,letterpaper]{revtex4-2}

\usepackage[colorlinks,linkcolor=blue,citecolor=blue,urlcolor=blue]{hyperref}

\usepackage{multirow}
\usepackage{subfigure}
\usepackage{color}
\usepackage{mathrsfs}
\usepackage{hyperref}
\usepackage[normalem]{ulem}
\usepackage{bm}

\usepackage{amssymb}   % for math
\usepackage{amsmath}
\usepackage{amsfonts, relsize, color}
\usepackage{graphics}
\usepackage{graphicx}
\usepackage{subfigure}
\usepackage{hyperref}
\usepackage{color}
\usepackage{comment}

\begin{document}

\title{Magnetic catalysis in weakly interacting hyperbolic Dirac materials}

\author{Bitan Roy}~\thanks{Corresponding author: bitan.roy@lehigh.edu}
\affiliation{Department of Physics, Lehigh University, Bethlehem, Pennsylvania, 18015, USA}

\date{\today}

\begin{abstract}
Due to the linearly vanishing density of states, emergent massless Dirac quasiparticles resulting from the free fermion motion in a family of two-dimensional half-filled bipartite hyperbolic lattices feature dynamic mass generation through quantum phase transitions only for sufficiently strong finite-range Coulomb repulsion. As such, strong nearest-neighbor Coulomb repulsion ($V$) favors the nucleation of a charge-density-wave (CDW) order with a staggered pattern of average fermionic density between two sublattices of bipartite hyperbolic lattices. Considering a collection of spinless fermions (for simplicity), here we show that application of strong external magnetic fields by virtue of producing a \emph{finite} density of states near the zero energy triggers the condensation of the CDW order even for \emph{infinitesimal} $V$. The proposed curved space magnetic catalysis mechanism is operative for uniform and inhomogeneous (bell-shaped) magnetic fields. We present scaling of the CDW order with the total flux enclosed by hyperbolic Dirac materials for a wide range of (especially subcritical) $V$.        
\end{abstract}

\maketitle

\section{Introduction}

Massless Dirac fermions living above one spatial dimension feature a vanishing density of states (DOS) near the half-filling or zero energy. As a result, dynamic mass generation of the bosonic order parameter field, composite of underlying fermionic degrees of freedom, through the spontaneous breaking of any discrete and/or continuous symmetries in Dirac materials occurs only at strong coupling in terms of quantum phase transitions. At the same time, the fermionic sector also becomes massive due to Yukawa-like interactions mediated by retarded short-range interactions  via the Anderson-Higgs mechanism. Such quantum phase transitions are typically triggered by the finite range components of Coulomb repulsion~\cite{ZinnJustin2021, Herbut2006, HJR2009, RoyDassarma2016}. Crystalline symmetry protection of the Dirac points in the Brillouin zone, where the filled valence and empty conduction bands touch each other, severely restricts the number of Euclidean lattices on which free-fermion motion gives birth to massless Dirac quasiparticles at half-filling. Graphene's honeycomb~\cite{Semenoff1984} and $\pi$-flux square~\cite{AffleckMarston1988} lattices are the lone members of this family, leaving aside some frustrated (such as kagome) lattices featuring Dirac fermions away from the half filling.

\begin{figure}[t!]
\includegraphics[width=1.00\linewidth]{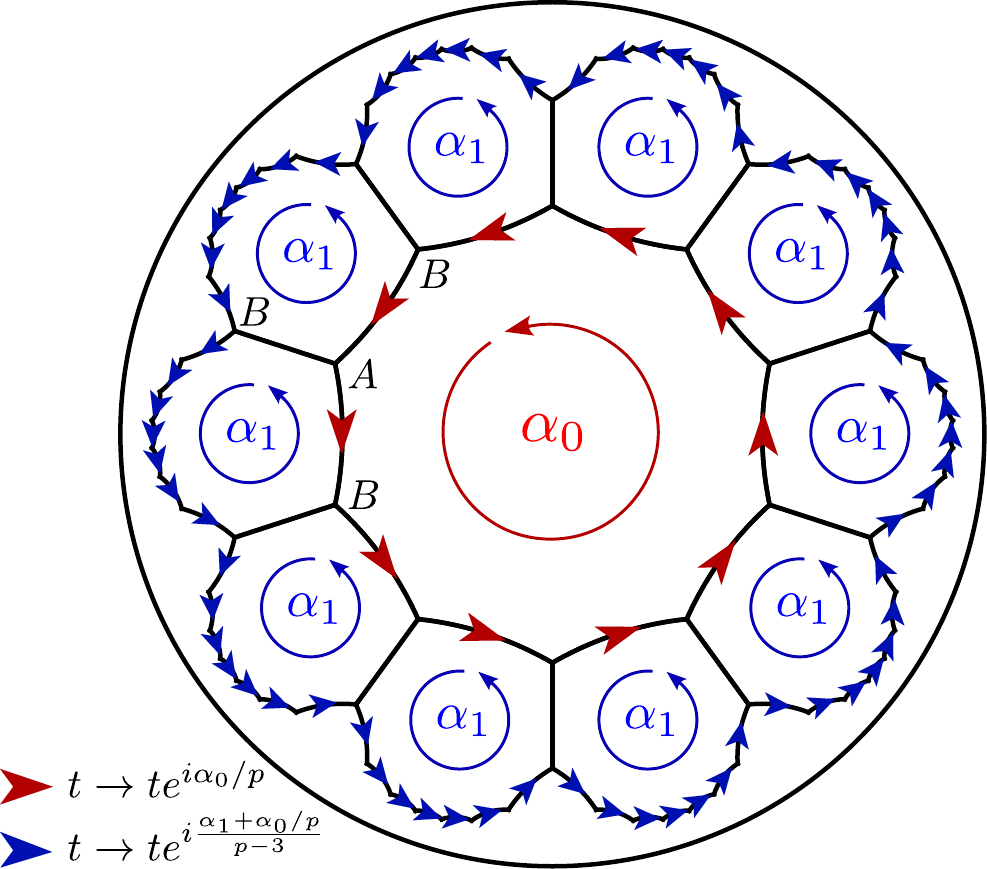}
\caption{Magnetic flux attachment to a $(p,q)=(10,3)$ hyperbolic lattice on a Poincar\'e disk through the Peierls substitution. The black lines represent a real and uniform hopping amplitude $t$ between the nearest-neighbor (NN) sites, belonging to the $A$ and $B$ sublattices. When accompanied by the red or blue arrows, the NN hopping amplitude becomes complex in its direction, with the phases shown in the legend. Then $\alpha_0$ ($\alpha_1$) amount of magnetic flux, measured in units of $\Phi_0/(2\pi)$, where $\Phi_0=h/e$ is the flux quantum, pierces through the plaquette(s) belonging to the zeroth (first) generation. This construction can be generalized to mimic an arbitrary flux profile to arbitrary generation on any hyperbolic lattice.         
}~\label{fig:Pierls}   
\end{figure}

Negatively curved hyperbolic quantum crystals, in this respect open a new direction. Throughout we characterize two-dimensional Euclidean and hyperbolic lattices, realized by repeated arrangements of polygons with $p$ arms of equal length ($p$-gons), each vertex of which is accompanied by $q$ equidistant nearest-neighbor (NN) sites, by a pair of integers $(p,q)$ (the Schl\"afli symbol). Then, a honeycomb [square] lattice is denoted by $(6,3)$ [$(4,4)$]. The geometric diversity of hyperbolic lattices, stemming from the inequality $(p-2)(q-2)>4$, gives rise to a variety of electronic band structures, captured by a simple NN tight-binding model for free fermions~\cite{KollarFitzpatrick2019, MaciejkoRayanSciAdv2021, BoettcherGorshkovPhysRevB2022, ChengSerafinPhysRevLett2022, MaciejkoRayan2022, AttarBoettcherPhysRevE2022, BzdusekMaciejkoPhysRevB2022, KienzleRayan2022, hyperbolicDiracRoy2023}. In particular, a family of hyperbolic lattices with $q=3$ harbors emergent massless Dirac fermions near the half-filling, when $p/2$ is an odd integer~\cite{hyperbolicDiracRoy2023}, hereafter called hyperbolic Dirac materials (HDMs). All hyperbolic lattices with odd $p/2$ are bipartite in nature, as then the NN sites belong to two sublattices, say, $A$ and $B$ (see Fig.~\ref{fig:Pierls}).

%%%%%%%%%%%%%%%%%%%%%%%%%%%%%%%%%%%%%%%%%%%%%%%%%%%%%%%%%%%%%%%%%%%%%%%%%%%%%%%%%%%%%%%%%%%%
%%%%%%%%%%%%%%%%%%%%%%%%%%%%%%%%%%%%%%%%%%%%%%%%%%%%%%%%%%%%%%%%%%%%%%%%%%%%%%%%%%%%%%%%%%%%
%%%%%%%%%%%%%%%%%%%%%%%%%%%%%%%%%%%%%%%%%%%%%%%%%%%%%%%%%%%%%%%%%%%%%%%%%%%%%%%%%%%%%%%%%%%%
%%%%%%%%%%%%%%%%%%%%%%%%%%%%%%%%%%%%%%%%%%%%%%%%%%%%%%%%%%%%%%%%%%%%%%%%%%%%%%%%%%%%%%%%%%%%
%%%%%%%%%%%%%%%%%%%%%%%%%%%%%%%%%%%%%%%%%%%%%%%%%%%%%%%%%%%%%%%%%%%%%%%%%%%%%%%%%%%%%%%%%%%%
%%%%%%%%%%%%%%%%%%%%%%%%%%%%%%%%%%%%%%%%%%%%%%%%%%%%%%%%%%%%%%%%%%%%%%%%%%%%%%%%%%%%%%%%%%%%
%%%%%%%%%%%%%%%%%%%%%%%%%%%%%%%%%%%%%%%%%%%%%%%%%%%%%%%%%%%%%%%%%%%%%%%%%%%%%%%%%%%%%%%%%%%%
%%%%%%%%%%%%%%%%%%%%%%%%%%%%%%%%%%%%%%%%%%%%%%%%%%%%%%%%%%%%%%%%%%%%%%%%%%%%%%%%%%%%%%%%%%%%
%%%%%%%%%%%%%%%%%%%%%%%%%%%%%%%%%%%%%%%%%%%%%%%%%%%%%%%%%%%%%%%%%%%%%%%%%%%%%%%%%%%%%%%%%%%%
%%%%%%%%%%%%%%%%%%%%%%%%%%%%%%%%%%%%%%%%%%%%%%%%%%%%%%%%%%%%%%%%%%%%%%%%%%%%%%%%%%%%%%%%%%%%

HDMs constitute an ideal platform to explore (theoretically and possibly experimentally) novel effects of electronic interactions among massless Dirac fermions living on a curved space. For example, strong NN Coulomb ($V$) and on-site Hubbard ($U$) repulsions support sublattice symmetry breaking charge-density-wave (CDW) and antiferromagnetic orders, respectively, displaying a staggered pattern of electronic density and magnetization between two sublattices. Furthermore, with increasing curvature or $p$, the critical strengths of $V$ and $U$ for these two orderings decrease monotonically, which are also smaller than their counterparts on a relativistic flatland (honeycomb lattice)~\cite{hyperbolicDiracRoy2023}. While this observation strongly suggests a fascinating phenomenon, a curvature induced quantum phase transition at weak coupling, no ordering develops for infinitesimal interactions in HDMs, due to the linearly \emph{vanishing} DOS near the half filling. This situation changes dramatically in the presence of external magnetic fields, when HDMs enclose finite magnetic flux, fostering magnetic catalysis of the CDW order at weak interactions. Next, we summarize the main findings of this work.  

%%%%%%%%%%%%%%%%%%%%%%%%%%%%%%%%%%%%%%%%%%%%%%%%%%%%%%%%%%%%%%%%%%%%%%%%%%%%%%%%%%%%%%%%%%%%
%%%%%%%%%%%%%%%%%%%%%%%%%%%%%%%%%%%%%%%%%%%%%%%%%%%%%%%%%%%%%%%%%%%%%%%%%%%%%%%%%%%%%%%%%%%%
%%%%%%%%%%%%%%%%%%%%%%%%%%%%%%%%%%%%%%%%%%%%%%%%%%%%%%%%%%%%%%%%%%%%%%%%%%%%%%%%%%%%%%%%%%%%
%%%%%%%%%%%%%%%%%%%%%%%%%%%%%%%%%%%%%%%%%%%%%%%%%%%%%%%%%%%%%%%%%%%%%%%%%%%%%%%%%%%%%%%%%%%%
%%%%%%%%%%%%%%%%%%%%%%%%%%%%%%%%%%%%%%%%%%%%%%%%%%%%%%%%%%%%%%%%%%%%%%%%%%%%%%%%%%%%%%%%%%%%
\begin{figure*}[t!]
\includegraphics[width=1.00\linewidth]{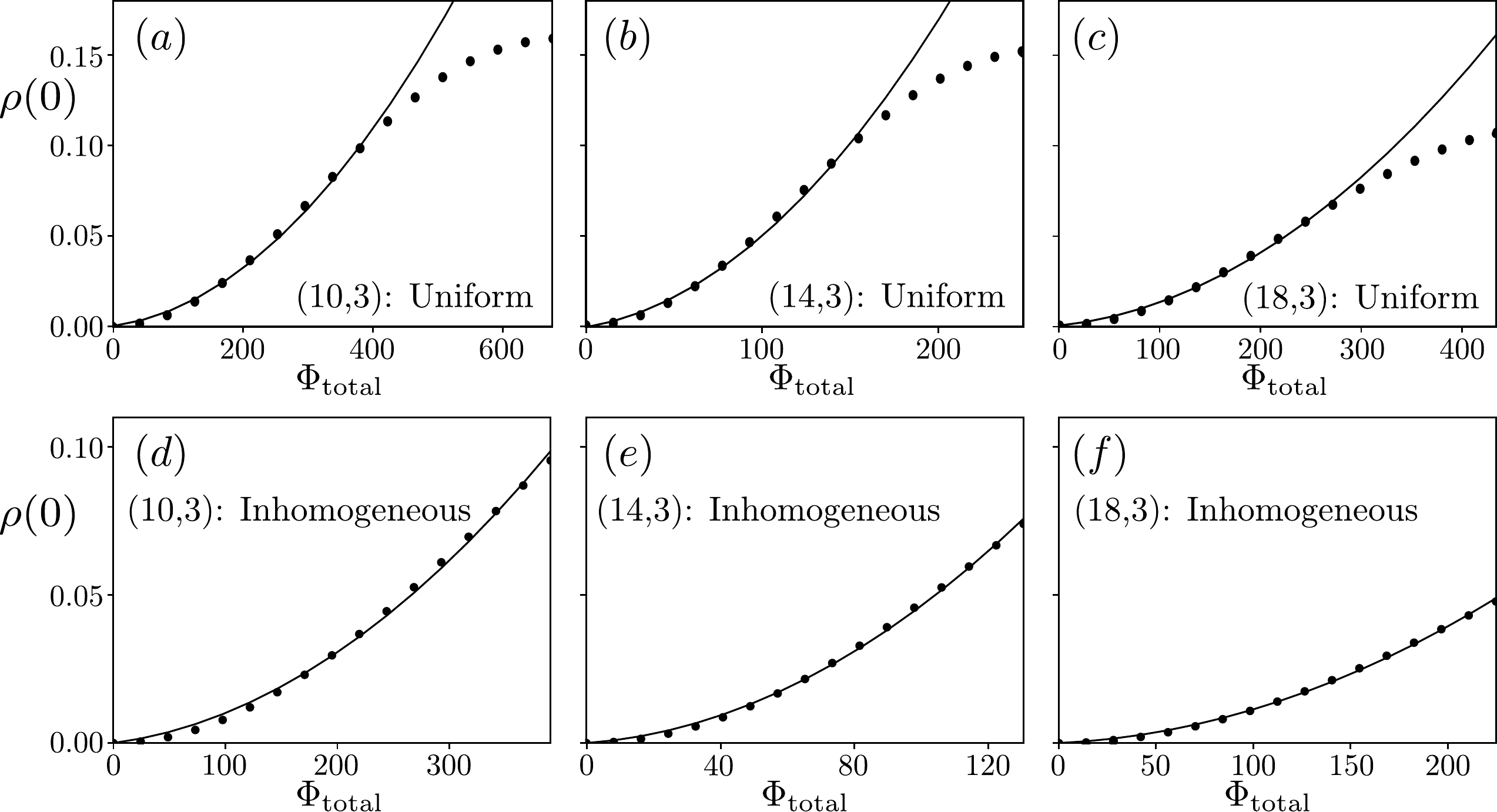}
\caption{Scaling of the density of states (DOS) at zero energy, defined as $\rho(0)=\rho_0(\Phi_{\rm total})-\rho_0(0)$, where $\rho_0(\Phi_{\rm total}) \; [\rho_0(0)]$ is the DOS with (without) magnetic field, with the total flux ($\Phi_{\rm total}$) enclosed by the (a) and (d) $(10,3)$, (b) and (e) $(14,3)$, and (c) and (f) $(18,3)$ hyperbolic lattices in the presence of (a)-(c) uniform and (d)-(f) bell-shaped inhomogeneous magnetic fields. For details see Sec.~\ref{subsec:fluxprofile}. As there are not many states near zero energy in a finite system, we use Gaussian smoothing to obtain a smooth curve, producing a background DOS at zero field $\rho_0(0)$, which we subtract to define $\rho(0)$ that increases with $\Phi_{\rm total}$, as detailed in Appendix~\ref{SMsec:DOSdetails}. However, the quadratic dependence of $\rho(0)$ on $\Phi_{\rm total}$ is obtained even without any Gaussian smoothing (see Fig.~\ref{fig:DOSappend}).  
}~\label{fig:DOS}   
\end{figure*}
%%%%%%%%%%%%%%%%%%%%%%%%%%%%%%%%%%%%%%%%%%%%%%%%%%%%%%%%%%%%%%%%%%%%%%%%%%%%%%%%%%%%%%%%%%%%
%%%%%%%%%%%%%%%%%%%%%%%%%%%%%%%%%%%%%%%%%%%%%%%%%%%%%%%%%%%%%%%%%%%%%%%%%%%%%%%%%%%%%%%%%%%%
%%%%%%%%%%%%%%%%%%%%%%%%%%%%%%%%%%%%%%%%%%%%%%%%%%%%%%%%%%%%%%%%%%%%%%%%%%%%%%%%%%%%%%%%%%%%
%%%%%%%%%%%%%%%%%%%%%%%%%%%%%%%%%%%%%%%%%%%%%%%%%%%%%%%%%%%%%%%%%%%%%%%%%%%%%%%%%%%%%%%%%%%%
%%%%%%%%%%%%%%%%%%%%%%%%%%%%%%%%%%%%%%%%%%%%%%%%%%%%%%%%%%%%%%%%%%%%%%%%%%%%%%%%%%%%%%%%%%%%

\subsection{Key results}

\begin{figure*}[t!]
\includegraphics[width=1.00\linewidth]{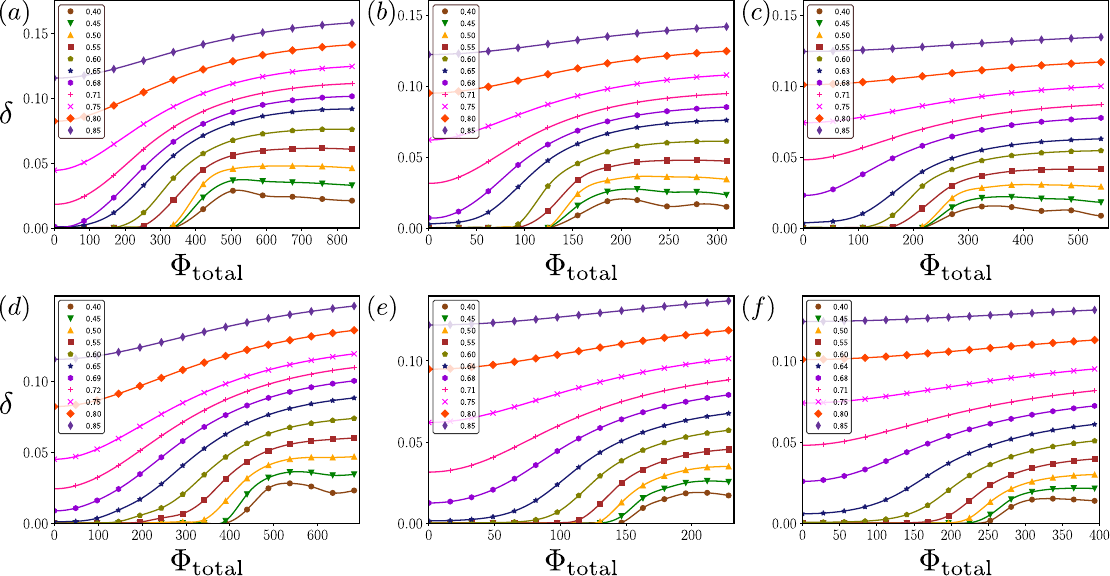}
\caption{Scaling of the charge-density-wave (CDW) order parameter $\delta$ [Eq.~(\ref{eq:CDWOP})], computed over the entire (a) and (d) (10,3), (b) and (e) (14,3), and (c) and (f) (18,3) hyperbolic lattices with the open boundary condition in the presence of (a)-(c) uniform and (d)-(f) bell-shaped inhomogeneous magnetic fields, with the total magnetic flux enclosed by the system ($\Phi_{\rm total}$) for a wide range of the NN Coulomb repulsion ($V$) among spinless fermions. For details see Sec.~\ref{subsec:fluxprofile}. The zero magnetic field critical NN Coulomb repulsion for the CDW ordering is $V_c \approx 0.69$, $0.67$, and $0.66$ in the (10,3), (14,3) and (18,3) hyperbolic lattices, respectively~\cite{hyperbolicDiracRoy2023}. Therefore, application of strong magnetic fields by virtue of developing a finite DOS near the zero energy (Fig.~\ref{fig:DOS}) catalyzes the formation of the CDW order for sufficiently weak NN Coulomb repulsion ($V \ll V_c$), yielding a correlated insulator near the half-filling (Fig.~\ref{fig:DOS_gap}).          
}~\label{fig:Scaling}   
\end{figure*}

We show that the application of strong magnetic fields (Fig.~\ref{fig:Pierls}) in HDMs creates a \emph{finite} DOS near zero energy or half filling that scales quadratically with the total flux enclosed by the system (see Fig.~\ref{fig:DOS}). Consequently, a sufficiently weak or subcritical NN Coulomb repulsion then nucleates a CDW order in these systems, a mechanism known as \emph{magnetic catalysis}~\cite{Miranskyetal1994, DunneHall1996, Khveshchenko2001, MiranskyQHE2006, HerbutCAF2007, RayaReyesInhomogeneous2010, RoyHerbutPhysRevB2011, RoyKennettDassarma2014}. These outcomes hold for a uniform magnetic field and a bell-shaped inhomogeneous magnetic field for which the field strength decreases monotonically from the center of the system toward its boundary (Fig.~\ref{fig:Scaling}). Formation of the CDW order splits the zero energy manifold and gives rise to a correlated insulator at half-filling (Fig.~\ref{fig:DOS_gap}). We arrive at these conclusions from numerical self-consistent solutions of the CDW order for (10,3), (14,3) and (18,3) HDMs with the open boundary condition in the presence of uniform and inhomogeneous magnetic fields for a wide range of subcritical NN Coulomb repulsion within the Hartree-Fock or mean-field approximation. Throughout we consider a collection of spinless fermions for simplicity. Additional numerical results supporting this mechanism are shown in Figs.~\ref{fig:DOSappend}-\ref{fig:cell}.

\subsection{Organization}

The rest of this paper is organized as follows. In Sec.~\ref{subsec:fluxprofile}, we specify the system details for HDMs and the profiles of the magnetic fields (uniform and inhomogeneous) adopted in this work. In Sec.~\ref{sec:freefermions}, we show how to capture the orbital effects of the external magnetic fields via Peierls substitution in the tight-binding model for free fermions and discuss the scaling of the DOS near zero energy with the magnetic flux. In Sec.~\ref{sec:NNCoulomb}, we introduce the NN Coulomb repulsion and arrive at the effective single-particle Hamiltonian after the Hartree decomposition. Section~\ref{sec:catalysis} is devoted to promoting the magnetic catalysis of CDW order in HDMs. We summarize the findings and present a discussion on related issues in Sec.~\ref{sec:summary}. Additional discussions and results are relegated to two Appendixes.

%%%%%%%%%%%%%%%%%%%%%%%%%%%%%%%%%%%%%%%%%%%%%%%%%%%%%%%%%%%%%%%%%%%%%%%%%%%%%%%%%%%%%%%%%%%%
%%%%%%%%%%%%%%%%%%%%%%%%%%%%%%%%%%%%%%%%%%%%%%%%%%%%%%%%%%%%%%%%%%%%%%%%%%%%%%%%%%%%%%%%%%%%
%%%%%%%%%%%%%%%%%%%%%%%%%%%%%%%%%%%%%%%%%%%%%%%%%%%%%%%%%%%%%%%%%%%%%%%%%%%%%%%%%%%%%%%%%%%%
%%%%%%%%%%%%%%%%%%%%%%%%%%%%%%%%%%%%%%%%%%%%%%%%%%%%%%%%%%%%%%%%%%%%%%%%%%%%%%%%%%%%%%%%%%%%
%%%%%%%%%%%%%%%%%%%%%%%%%%%%%%%%%%%%%%%%%%%%%%%%%%%%%%%%%%%%%%%%%%%%%%%%%%%%%%%%%%%%%%%%%%%%
%%%%%%%%%%%%%%%%%%%%%%%%%%%%%%%%%%%%%%%%%%%%%%%%%%%%%%%%%%%%%%%%%%%%%%%%%%%%%%%%%%%%%%%%%%%%
%%%%%%%%%%%%%%%%%%%%%%%%%%%%%%%%%%%%%%%%%%%%%%%%%%%%%%%%%%%%%%%%%%%%%%%%%%%%%%%%%%%%%%%%%%%%
%%%%%%%%%%%%%%%%%%%%%%%%%%%%%%%%%%%%%%%%%%%%%%%%%%%%%%%%%%%%%%%%%%%%%%%%%%%%%%%%%%%%%%%%%%%%
%%%%%%%%%%%%%%%%%%%%%%%%%%%%%%%%%%%%%%%%%%%%%%%%%%%%%%%%%%%%%%%%%%%%%%%%%%%%%%%%%%%%%%%%%%%%
%%%%%%%%%%%%%%%%%%%%%%%%%%%%%%%%%%%%%%%%%%%%%%%%%%%%%%%%%%%%%%%%%%%%%%%%%%%%%%%%%%%%%%%%%%%%

\section{System specifications and magnetic flux profiles}~\label{subsec:fluxprofile}

First, we outline the details of the systems and the magnetic flux profiles considered in this work. All the numerical analyses are performed on a third generation (10,3), and second generation (14,3) and (18,3) hyperbolic lattices, respectively containing 2880 (421), 1694 (155) and 4050 (271) lattice sites ($p$-gons or plaquettes), respectively. The center plaquette corresponds to the zeroth generation and each successive layers of plaquettes constitutes the progressively next generation of the hyperbolic lattice. When the field is uniform, equal flux ($\alpha_0$) threads all the plaquettes. A bell-shaped inhomogeneous magnetic field is modeled by threading $\alpha_0$, $0.85 \alpha_0$, $0.70 \alpha_0$, and $0.55 \alpha_0$ amount of magnetic fluxes through all the plaquettes belonging to the zeroth, first, second, and third generations of the (10,3) hyperbolic lattice, respectively. On (14,3) and (18,3) hyperbolic lattices $\alpha_0$, $0.75 \alpha_0$, and $0.50 \alpha_0$ magnetic flux pierces through each plaquette belonging to the zeroth, first, and second generations, respectively. We measure the magnetic flux in units of $\Phi_0/(2\pi)$, where $\Phi_0=h/e$ is the magnetic flux quantum.

Although in this work we consider (10,3), (14,3), and (18,3) hyperbolic lattices with finite number of sites in the system with the open boundary condition, such tessellations, always respecting the inequality $(p-2)(q-2)>4$, can be accommodated only on a negatively curved space, not on the flat Euclidean plane. Thus, all the outcomes (such as the quadratic dependence of the DOS near zero-energy with total flux enclosed by the system, which discussed in the next section) manifest the underlying curved hyperbolic space on which the lattice sites reside.

\section{Free fermions and zero modes}~\label{sec:freefermions}

We begin the discussion with a tight-binding Hamiltonian for free fermions $H_0$, allowed to hop only between the NN sites of the hyperbolic lattices, subject to external magnetic fields. For spinless fermions, only the orbital effect of the external magnetic field is pertinent, captured by the Peierls substitution~\cite{Peierls1933}. Then 
\begin{equation}~\label{eq:TBPeierls}
H_0 =-\sum_{j \in A} \sum_{k \in B}{}^{'} t_{jk}\; c^\dagger_j \; \exp[i \alpha_{jk}] \; c_k + {\rm H.c.}\;,  
\end{equation}    
where $i=\sqrt{-1}$, $c^\dagger_j$ ($c_j$) is the fermionic creation (annihilation) operator on the $j$th site, and the prime symbol restricts the summation within the NN sites. The sublattice labeling of the sites is arbitrary, manifesting an Isinglike sublattice exchange symmetry. The spin independent NN hopping amplitude $t_{jk}$ is assumed to be constant $t$, which we set to be unity. As shown in Fig.~\ref{fig:Pierls}, this prescription can be employed on any hyperbolic lattice to mimic any spatial profile of the magnetic field. As such uniform magnetic fields can display many peculiar phenomena on hyperbolic lattices~\cite{comtethouston1985, ludewigThiang2021, IkedaAokiJournalofPhysics2021, StegmaierUpretiPhysRevLett2022, MosseriVogelerPhysRevB2022}, such as the Hofstadter butterfly, devoid of fractal structures~\cite{StegmaierUpretiPhysRevLett2022}.

%%%%%%%%%%%%%%%%%%%%%%%%%%%%%%%%%%%%%%%%%%%%%%%%%%%%%%%%%%%%%%%%%%%%%%%%%%%%%%%%%%%%%%%%%%%%
%%%%%%%%%%%%%%%%%%%%%%%%%%%%%%%%%%%%%%%%%%%%%%%%%%%%%%%%%%%%%%%%%%%%%%%%%%%%%%%%%%%%%%%%%%%%
%%%%%%%%%%%%%%%%%%%%%%%%%%%%%%%%%%%%%%%%%%%%%%%%%%%%%%%%%%%%%%%%%%%%%%%%%%%%%%%%%%%%%%%%%%%%
%%%%%%%%%%%%%%%%%%%%%%%%%%%%%%%%%%%%%%%%%%%%%%%%%%%%%%%%%%%%%%%%%%%%%%%%%%%%%%%%%%%%%%%%%%%%
%%%%%%%%%%%%%%%%%%%%%%%%%%%%%%%%%%%%%%%%%%%%%%%%%%%%%%%%%%%%%%%%%%%%%%%%%%%%%%%%%%%%%%%%%%%%
\begin{figure*}[t!]
\includegraphics[width=1.00\linewidth]{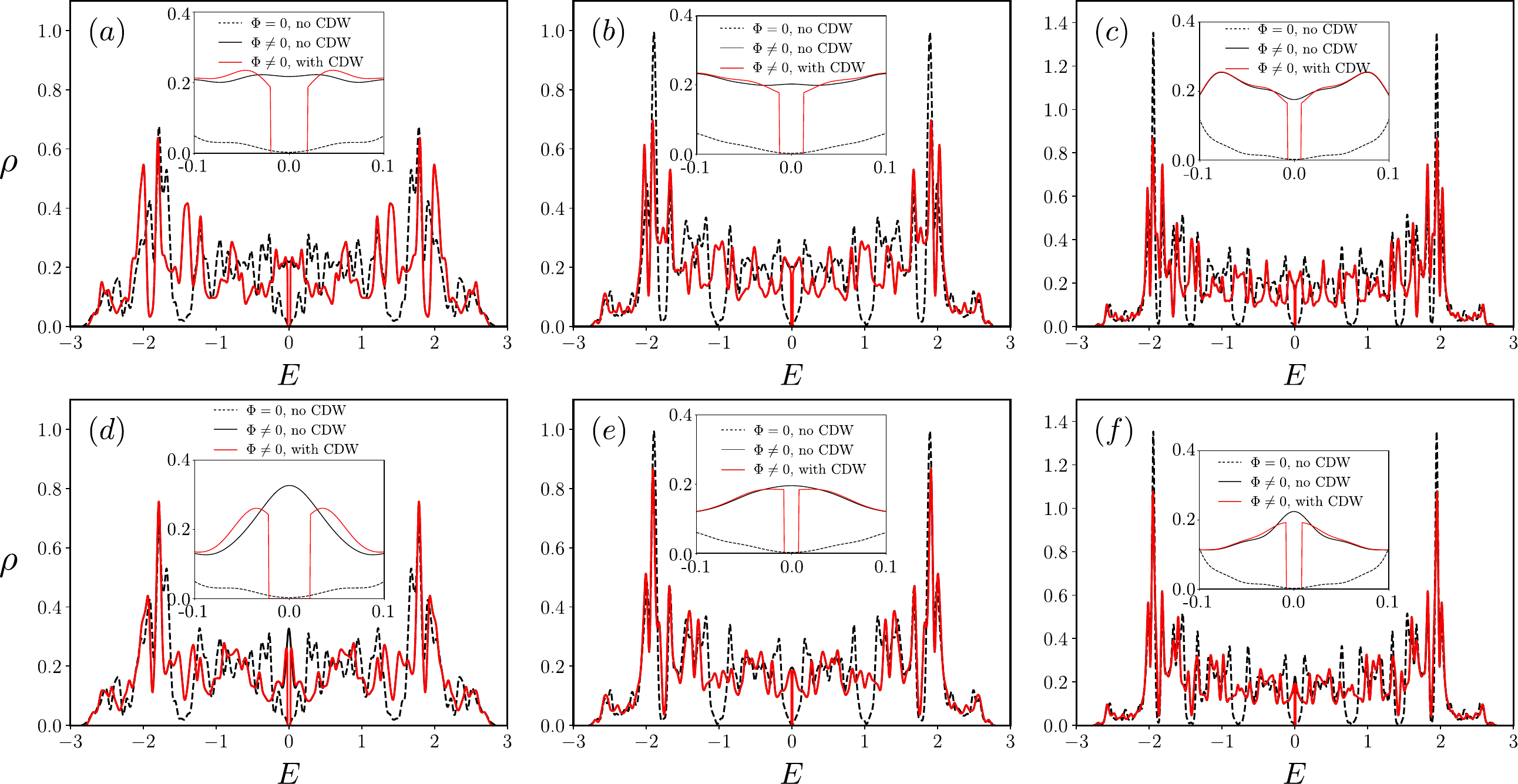}
\caption{Density of states ($\rho$) vs energy ($E$) in the absence (black dashed lines) and presence (black solid lines) of magnetic flux without the charge-density-wave order in the (a) and (d) (10,3), (b) and (e) (14,3), and (c) and (f) (18,3) hyperbolic lattices in the presence of (a)-(c) uniform and (d)-(f) inhomogeneous magnetic flux. Red solid lines correspond to $\rho$ with the field-induced charge-density-wave order for the nearest-neighbor Coulomb repulsion $V=0.4$ ($\ll V_c$). The total magnetic flux threading the system $\Phi_{\rm total}$ is (a) 842, (b) 310, (c) 542, (d) 488, (e) 163, and (f) 281 [in units of $\Phi_0/(2\pi)$]. Insets depict the formation of correlated insulators with a sharp spectral gap near zero energy for subcritical $V$ via the curved space magnetic catalysis mechanism. Recall that $V_c \approx 0.69$, $0.67$ and $0.66$ in the (10,3), (14,3) and (18,3) hyperbolic lattices, respectively~\cite{hyperbolicDiracRoy2023}.    
}~\label{fig:DOS_gap}   
\end{figure*}
%%%%%%%%%%%%%%%%%%%%%%%%%%%%%%%%%%%%%%%%%%%%%%%%%%%%%%%%%%%%%%%%%%%%%%%%%%%%%%%%%%%%%%%%%%%%
%%%%%%%%%%%%%%%%%%%%%%%%%%%%%%%%%%%%%%%%%%%%%%%%%%%%%%%%%%%%%%%%%%%%%%%%%%%%%%%%%%%%%%%%%%%%
%%%%%%%%%%%%%%%%%%%%%%%%%%%%%%%%%%%%%%%%%%%%%%%%%%%%%%%%%%%%%%%%%%%%%%%%%%%%%%%%%%%%%%%%%%%%
%%%%%%%%%%%%%%%%%%%%%%%%%%%%%%%%%%%%%%%%%%%%%%%%%%%%%%%%%%%%%%%%%%%%%%%%%%%%%%%%%%%%%%%%%%%%
%%%%%%%%%%%%%%%%%%%%%%%%%%%%%%%%%%%%%%%%%%%%%%%%%%%%%%%%%%%%%%%%%%%%%%%%%%%%%%%%%%%%%%%%%%%%

We focus near the zero energy of half-filled HDMs, subject to magnetic fields. They possess a particle-hole symmetry about the zero energy and keep all the negative (positive) energy states occupied (empty). The average fermionic density at each site is then $1/2$, manifesting the sublattice exchange symmetry. In Euclidean Dirac systems, application of external magnetic fields produces zero energy states, the number of which is proportional to the total magnetic flux enclosed by the system ($\Phi_{\rm total}$), irrespective of the magnetic field profile: Aharonov-Casher index theorem~\cite{AharonovCasher1979}. This index theorem has been verified from honeycomb lattice-based exact numerical diagonalization~\cite{RoyHerbutPhysRevB2011} and possibly also extends to HDMs~\cite{InahamaShirai2003, MineNomura2012}.

Here we compute the DOS near the zero energy $\rho(0)$ from exact numerical diagonalization of $H_0$ [Eq.~(\ref{eq:TBPeierls}); and See Fig.~\ref{fig:DOS}]. The absence of infinitely degenerate zeroth and other Landau levels in HDMs even when the magnetic field is uniform possibly stems from their nontrivial spatial curvature. In the small flux regime $\rho(0)$ scales \emph{quadratically} with $\Phi_{\rm total}$ for both uniform and bell-shaped inhomogeneous magnetic fields. In the large flux regime, the scaling of $\rho(0)$ deviates from the quadratic dependence on $\Phi_{\rm total}$, when the magnetic length becomes comparable to the lattice constant. These observations qualitatively conform to the Aharonov-Casher index theorem, extended to HDMs, possibly indicating a topological protection of the magnetic-field-induced near-zero-energy modes. Most crucially, when immersed in magnetic fields HDMs always supports a finite DOS near zero energy. The appearance of a finite number of states near half filling can be conducive to the nucleation of ordered phases even for sufficiently weak finite-range Coulomb repulsion, which we discuss next.

Notice that the quadratic dependence of $\rho(0)$ on $\Phi_{\rm total}$, depicted in Fig.~\ref{fig:DOS}, is obtained after a Gaussian smoothing, as detailed in Appendix~\ref{SMsec:DOSdetails}. However, such a scaling holds even without the Gaussian smoothing (see Fig.~\ref{fig:DOSappend}). Notice that the quadratic dependence of $\rho(0)$ on $\Phi_{\rm total}$ in hyperbolic Dirac systems is distinct from the linear dependence of $\rho(0)$ on $\Phi_{\rm total}$ for Dirac fermions, living on the flat space~\cite{AharonovCasher1979}. Such a scaling unfolds the imprint of the underlying negatively curved hyperbolic plane, decorated by the sites of the hyperbolic lattices.

\section{Nearest-neighbor interaction}~\label{sec:NNCoulomb}

The zero energy manifold in HDMs, featuring a finite DOS and sourced by external magnetic fields, can be split by weak enough NN Coulomb repulsion $V$ through a dynamic breaking of the sublattice exchange symmetry. For a collection of $N$ spinless fermions, the corresponding  Hamiltonian reads
\begin{equation}
H_V = H_0 + \frac{V}{2} \sum_{\langle j, k \rangle} n_j n_k - \mu N.
\end{equation}
Here $n_j=c^\dagger_j c_j$ is the fermionic density on site $j$, $\mu$ is the chemical potential in the half-filled system, and $\langle \cdots \rangle$ restricts the summation to the NN sites.

Hartree decomposition of the quartic term then leads to the following effective single-particle Hamiltonian~\cite{RoyHerbutPhysRevB2011, RoySau2014, RoyPhysRevB2017} 
\begin{equation}
H^{\rm Har}_V= H_0 + V \sum_{\langle j,k \rangle} \bigg[ \langle n_{A,j} \rangle n_{B, k} + \langle n_{B,j} \rangle n_{A, k} \bigg] - \mu N,
\end{equation}
where $\langle n_{A} \rangle$ ($\langle n_{B} \rangle$) corresponds to the site dependent self-consistent average fermionic density on the sublattice $A$ ($B$). We measure the densities relative to the uniform density at half-filling according to $\langle n_{A,j}\rangle = 1/2 + \delta_{A,j}$ and $\langle n_{B,j}\rangle = 1/2 - \delta_{B,j}$. To maintain the system at the half-filling, we choose $\mu=V/2$ and ensure that $\sum_{j} \left( \delta_{A,j} -\delta_{B,j} \right) =0$. The positive-definite quantities $\delta_A$ and $\delta_B$ yield the CDW order parameter in the whole system, defined as 
\begin{equation}~\label{eq:CDWOP}
\delta=\frac{1}{N} \bigg( \sum_{j} \delta_{A,j} + \sum_{j} \delta_{B,j} \bigg). 
\end{equation}
We numerically compute $\delta_{A,j}$ and $\delta_{B,j}$, and subsequently $\delta$ in the entire system with the open boundary condition for a wide range of $V$, especially for its subcritical strengths, in the presence of uniform and inhomogeneous magnetic fields of varying $\Phi_{\rm total}$. For details on various systems and flux profiles, see Sec.~\ref{subsec:fluxprofile}. The results are shown in Fig.~\ref{fig:Scaling}, which we discuss next.

\section{Magnetic catalysis of CDW}~\label{sec:catalysis}

The zero magnetic field critical strengths of the NN Coulomb repulsion (obtained within the Hartree approximation) for the CDW ordering are $V_c=0.69$, $0.67$, and $0.66$ in the (10,3), (14,3) and (18,3) hyperbolic lattices of the specified generation number, respectively, as mentioned in Sec.~\ref{subsec:fluxprofile}~\cite{hyperbolicDiracRoy2023}. As shown in Fig.~\ref{fig:Scaling}, in the presence of finite magnetic fluxes $\delta_A$ and $\delta_B$, concomitantly, $\delta$ becomes finite in the entire system even when $V \ll V_c$ in these three HDMs. The spatial profile of the CDW order for various generations of the HDMs in the presence of magnetic flux is shown in Fig.~\ref{fig:gendependence} and discussed in Appendix~\ref{SMsec:additionalnumerics}. This outcome holds for uniform as well as bell-shaped inhomogeneous magnetic field, piercing the system. As external magnetic fields produce finite $\rho(0)$ (Fig.~\ref{fig:DOS}), a sizable condensation of the CDW order parameter ($\delta$) takes place only in the large flux limit when $V \ll V_c$. Such an outcome can be appreciated from its BCS like scaling, 
\begin{equation}
\delta = a \exp \left( -b/[\Phi_{\rm total}-\Phi_{\rm th}] \right)
\end{equation}  
for $\Phi_{\rm total}>\Phi_{\rm th}$ and any fixed $V<V_c$, resulting from the finite DOS near the zero energy~\cite{roysaucatalysis}. Here $a$, $b$, and $\Phi_{\rm th}$ are fitting parameters. For example, with uniform [inhomogeneous] magnetic fields $(a,b,\Phi_{\rm th}) \approx$ (0.071, 70, 260) [(0.090, 175, 215)], (0.051, 17, 111) [(0.075, 70, 85)] and (0.050, 50, 180) [(0.050, 50, 180)] in the (10,3), (14,3), and (18,3) HDMs for $V=0.55$, respectively. A nonzero $\Phi_{\rm th}$ stems from finite size effect for weak $V$, which was also previously noticed for the weak coupling antiferromagnetism driven by Hubbard repulsion in Bernal stacked bilayer graphene~\cite{AssaadBBLG2012} and the onset of a diffusive metal for weak enough disorder in double-Weyl semimetals~\cite{BeraSauRoyDWSM2016}. Nevertheless, a finite CDW order develops in the entire system for any subcritical strength of $V$ and for any $\Phi_{\rm total}$, supporting the magnetic catalysis mechanism of this order in HDMs. As the strength of subcritical $V$ increases, $\delta$ becomes appreciable for small $\Phi_{\rm total}$ and $\Phi_{\rm th}$ decreases monotonically. To unambiguously establish the magnetic catalysis mechanism of HDMs, we compute the difference between finite and zero magnetic field CDW order parameter. For all choices of $V<V_c$ and $\Phi_{\rm total}$, this quantity is positive definite, confirming that the formation of the CDW order for subcritical NN Coulomb interaction is solely due to the magnetic fields. See Fig.~\ref{fig:cell} and the discussion in Appendix~\ref{SMsec:additionalnumerics}. For $V>V_c$, $\delta$ is finite even in the absence of magnetic field, which then increases as $\delta \sim \Phi^2_{\rm total}$ in the small flux regime.

%%%%%%%%%%%%%%%%%%%%%%%%%%%%%%%%%%%%%%%%%%%%%%%%%%%%%%%%%%%%%%%%%%%%%%%%%%%%%%%%%%%%%%%%%%%%
%%%%%%%%%%%%%%%%%%%%%%%%%%%%%%%%%%%%%%%%%%%%%%%%%%%%%%%%%%%%%%%%%%%%%%%%%%%%%%%%%%%%%%%%%%%%
%%%%%%%%%%%%%%%%%%%%%%%%%%%%%%%%%%%%%%%%%%%%%%%%%%%%%%%%%%%%%%%%%%%%%%%%%%%%%%%%%%%%%%%%%%%%
%%%%%%%%%%%%%%%%%%%%%%%%%%%%%%%%%%%%%%%%%%%%%%%%%%%%%%%%%%%%%%%%%%%%%%%%%%%%%%%%%%%%%%%%%%%%
%%%%%%%%%%%%%%%%%%%%%%%%%%%%%%%%%%%%%%%%%%%%%%%%%%%%%%%%%%%%%%%%%%%%%%%%%%%%%%%%%%%%%%%%%%%%
\begin{figure}[t!]
\includegraphics[width=1.00\linewidth]{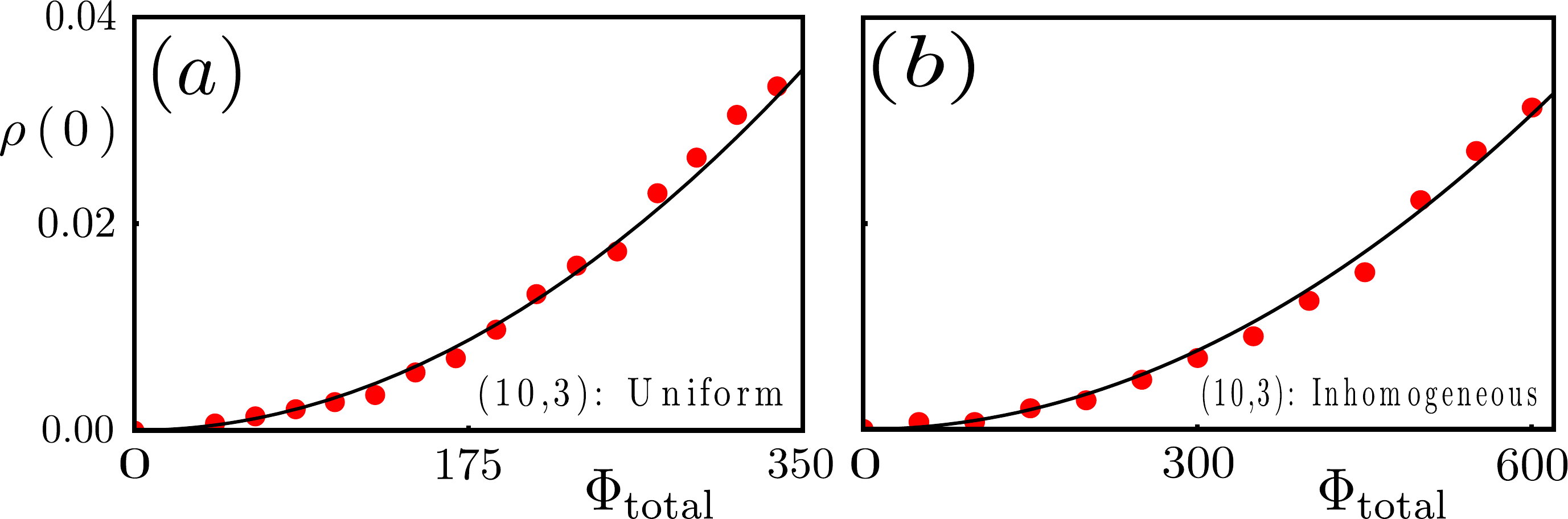}
\caption{Scaling of the DOS at zero energy, defined as $\rho(0)=\rho_0(\Phi_{\rm total})-\rho_0(0)$, where $\rho_0(\Phi_{\rm total}) \; [\rho_0(0)]$ is the DOS with (without) magnetic field, with the total flux ($\Phi_{\rm total}$) enclosed by the $(10,3)$ hyperbolic lattice in the presence of (a) uniform and (b) bell-shaped inhomogeneous magnetic fields. Here, we compute the DOS within the energy window $\Delta E=0.11$ and do not employ any Gaussian smoothing. Still, we find $\rho(0) \sim \Phi^2_{\rm total}$. For details see Sec.~\ref{subsec:fluxprofile} and Appendix~\ref{SMsec:DOSdetails}. 
}~\label{fig:DOSappend}   
\end{figure}
%%%%%%%%%%%%%%%%%%%%%%%%%%%%%%%%%%%%%%%%%%%%%%%%%%%%%%%%%%%%%%%%%%%%%%%%%%%%%%%%%%%%%%%%%%%%
%%%%%%%%%%%%%%%%%%%%%%%%%%%%%%%%%%%%%%%%%%%%%%%%%%%%%%%%%%%%%%%%%%%%%%%%%%%%%%%%%%%%%%%%%%%%
%%%%%%%%%%%%%%%%%%%%%%%%%%%%%%%%%%%%%%%%%%%%%%%%%%%%%%%%%%%%%%%%%%%%%%%%%%%%%%%%%%%%%%%%%%%%
%%%%%%%%%%%%%%%%%%%%%%%%%%%%%%%%%%%%%%%%%%%%%%%%%%%%%%%%%%%%%%%%%%%%%%%%%%%%%%%%%%%%%%%%%%%%
%%%%%%%%%%%%%%%%%%%%%%%%%%%%%%%%%%%%%%%%%%%%%%%%%%%%%%%%%%%%%%%%%%%%%%%%%%%%%%%%%%%%%%%%%%%%

When $\delta$ is finite, HDMs become a correlated insulator at half-filling by spontaneously breaking the sublattice exchange symmetry through the curved space magnetic catalysis mechanism for $V<V_c$. To demonstrate this outcome we define a two-component superspinor $\Psi^\top=(c_A, c_B)$, where $c_A$ ($c_B$) is an $N$-dimensional spinor constituted by the annihilation operators on the sites of sublattice $A$ ($B$). In this basis, the tight-binding Hamiltonian in the presence of magnetic fields and the Hamiltonian with the CDW order parameter are
\begin{equation}~\label{eq:effectivehamil}
\hat{h}_0= \left( \begin{array}{cc}
{\boldsymbol 0} & {\bf t} \\
{\bf t}^\dagger & {\bf 0}
\end{array}
\right)
\:\: \text{and} \:\:
\hat{h}_{\rm CDW} =\left( \begin{array}{cc}
{\boldsymbol \Delta} & {\boldsymbol 0} \\
{\boldsymbol 0} & -{\boldsymbol \Delta}
\end{array} 
\right),
\end{equation}    
 respectively, where ${\boldsymbol 0}$ is an $N$-dimensional null matrix, ${\bf t}$ and ${\bf t}^\dagger$ are the intersublattice hopping matrices with the Peierls phase factors, and $\pm {\boldsymbol \Delta}$ are $N$-dimensional diagonal matrices, whose entries are the self-consistent solutions of $\delta_A$ and $\delta_B$ at various sites of the system, respectively. As $\hat{h}_0$ and $\hat{h}_{\rm CDW}$ mutually \emph{anticommute}, the CDW order acts like a mass for gapless fermions on HDMs, subject to external magnetic fields. Thus the spontaneous nucleation of the CDW order [Fig.~\ref{fig:Scaling}] causes insulation near the half-filling even for small $V$, as shown in Fig.~\ref{fig:DOS_gap} for both uniform and inhomogeneous magnetic fields on the (10,3), (14,3), and (18,3) hyperbolic lattices. 

%%%%%%%%%%%%%%%%%%%%%%%%%%%%%%%%%%%%%%%%%%%%%%%%%%%%%%%%%
%%%%%%%%%%%%%%%%%%%%%%%%%%%%%%%%%%%%%%%%%%%%%%%%%%%%%%%%%
%%%%%%%%%%%%%%%%%%%%%%%%%%%%%%%%%%%%%%%%%%%%%%%%%%%%%%%%%
%%%%%%%%%%%%%%%%%%%%%%%%%%%%%%%%%%%%%%%%%%%%%%%%%%%%%%%%%
%%%%%%%%%%%%%%%%%%%%%%%%%%%%%%%%%%%%%%%%%%%%%%%%%%%%%%%%%
\begin{figure*}[t!]
\includegraphics[width=1.00\linewidth]{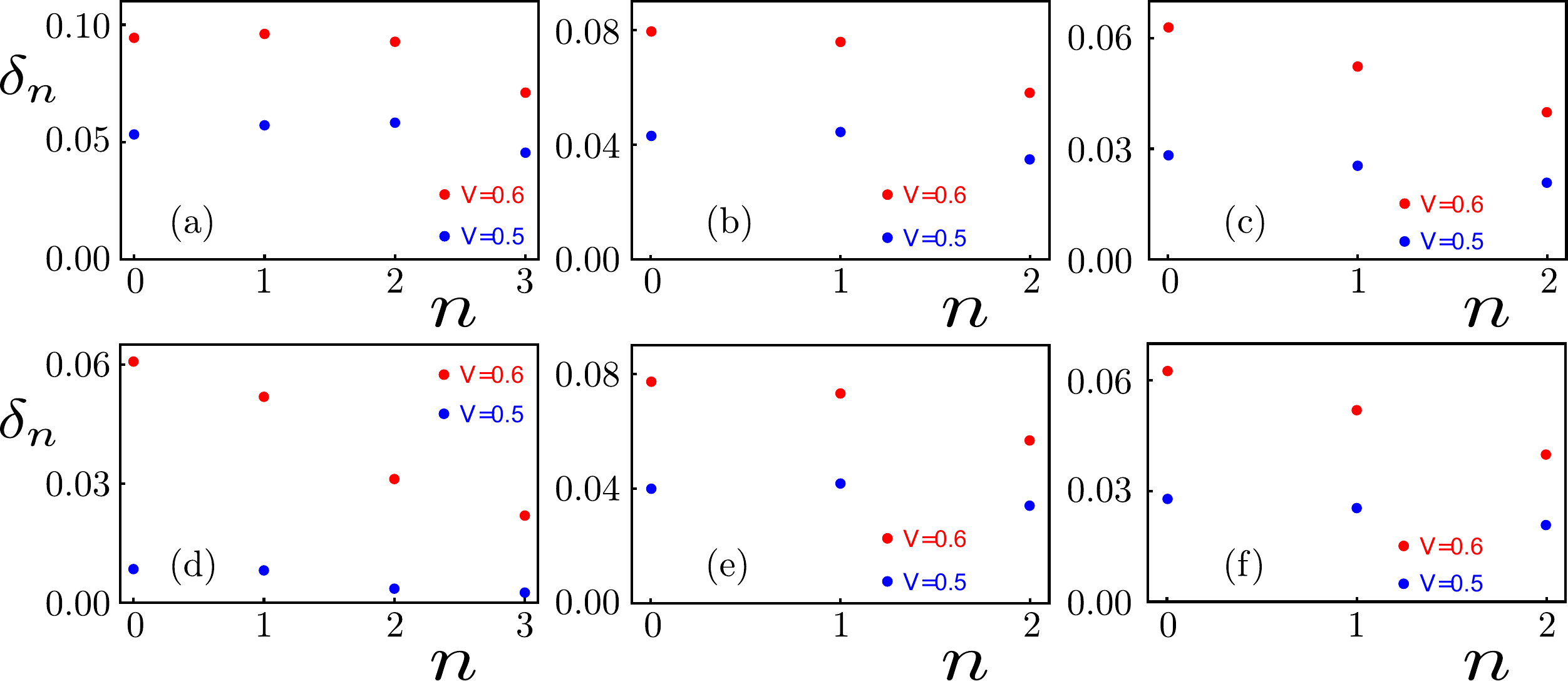}
\caption{The local CDW order parameter ($\delta_n$) [see Eq.~\eqref{eq:CDWOPLocal}], developed on the sites, belonging to the $n$th generation of the (a) and (d) (10,3), (b) and (e) (14,3), and (c) and (f) (18,3) hyperbolic lattices in the presence of (a)-(c) uniform and (d)-(f) inhomogeneous magnetic flux for the nearest-neighbor Coulomb repulsion $V=0.6$ and $V=0.5$. The total flux enclosed by the system $\Phi_{\rm total}$  is (a) $673.6$, (b) $248.0$, (c) $271.0$, (d) $683.2$, (e) $244.5$, and (f) $281.0$ [in units of $\Phi_0/(2\pi)$]. For specifications of these systems and the magnetic flux threading systems see Sec.~\ref{subsec:fluxprofile}. Note that $V=0.6,0.5<V_c$ (critical nearest-neighbor Coulomb repulsion for the CDW order in the absence of any magnetic field) in all the systems, showing that the CDW order develops in the entire system via the magnetic catalysis mechanism. Recall that $V_c \approx 0.69$, $0.67$ and $0.66$ in the (10,3), (14,3), and (18,3) hyperbolic lattices, respectively~\cite{hyperbolicDiracRoy2023}. For details see Appendix~\ref{SMsec:additionalnumerics}. 
}~\label{fig:gendependence}
\end{figure*}
%%%%%%%%%%%%%%%%%%%%%%%%%%%%%%%%%%%%%%%%%%%%%%%%%%%%%%%%%
%%%%%%%%%%%%%%%%%%%%%%%%%%%%%%%%%%%%%%%%%%%%%%%%%%%%%%%%%
%%%%%%%%%%%%%%%%%%%%%%%%%%%%%%%%%%%%%%%%%%%%%%%%%%%%%%%%%
%%%%%%%%%%%%%%%%%%%%%%%%%%%%%%%%%%%%%%%%%%%%%%%%%%%%%%%%%
%%%%%%%%%%%%%%%%%%%%%%%%%%%%%%%%%%%%%%%%%%%%%%%%%%%%%%%%%

\section{Summary and discussion}~\label{sec:summary}

We showed that the application of external uniform or inhomogeneous magnetic fields in HDMs, featuring a linearly vanishing DOS in the pristine condition, gives rise to a finite DOS near the zero energy which increases quadratically with the total flux enclosed by the system. Then the curved space magnetic catalysis becomes operative in the entire HDM family, through which a CDW ordering nucleates even for subcritical NN Coulomb repulsion among (spinless) fermions. Inclusion of the spin degrees of freedom brings the on-site Hubbard repulsion onto the stage, which is conducive to the formation of an antiferromagntic order if it is sufficiently strong~\cite{hyperbolicDiracRoy2023}. When a magnetic field penetrates HDMs an easy-plane (perpendicular to the field direction) antiferromagnet, accompanied by a Zeeman coupling induced ferromagnetic order in the field direction is expected to emerge as the ground state for a subcritical strength of the Hubbard repulsion~\cite{HerbutCAF2007, RoyKennettDassarma2014}. A detailed study of this competition is left for a future investigation. It would also be fascinating to develop a field-theoretic description of this phenomenon in terms of the Gross-Neveu model~\cite{GrossNeveu} for Dirac fermions on curved space, subject to magnetic fields.

%%%%%%%%%%%%%%%%%%%%%%%%%%%%%%%%%%%%%%%%%%%%%%%%%%%%%%%%%
%%%%%%%%%%%%%%%%%%%%%%%%%%%%%%%%%%%%%%%%%%%%%%%%%%%%%%%%%
%%%%%%%%%%%%%%%%%%%%%%%%%%%%%%%%%%%%%%%%%%%%%%%%%%%%%%%%%
%%%%%%%%%%%%%%%%%%%%%%%%%%%%%%%%%%%%%%%%%%%%%%%%%%%%%%%%%
%%%%%%%%%%%%%%%%%%%%%%%%%%%%%%%%%%%%%%%%%%%%%%%%%%%%%%%%%
\begin{figure*}[t!]
\includegraphics[width=1.00\linewidth]{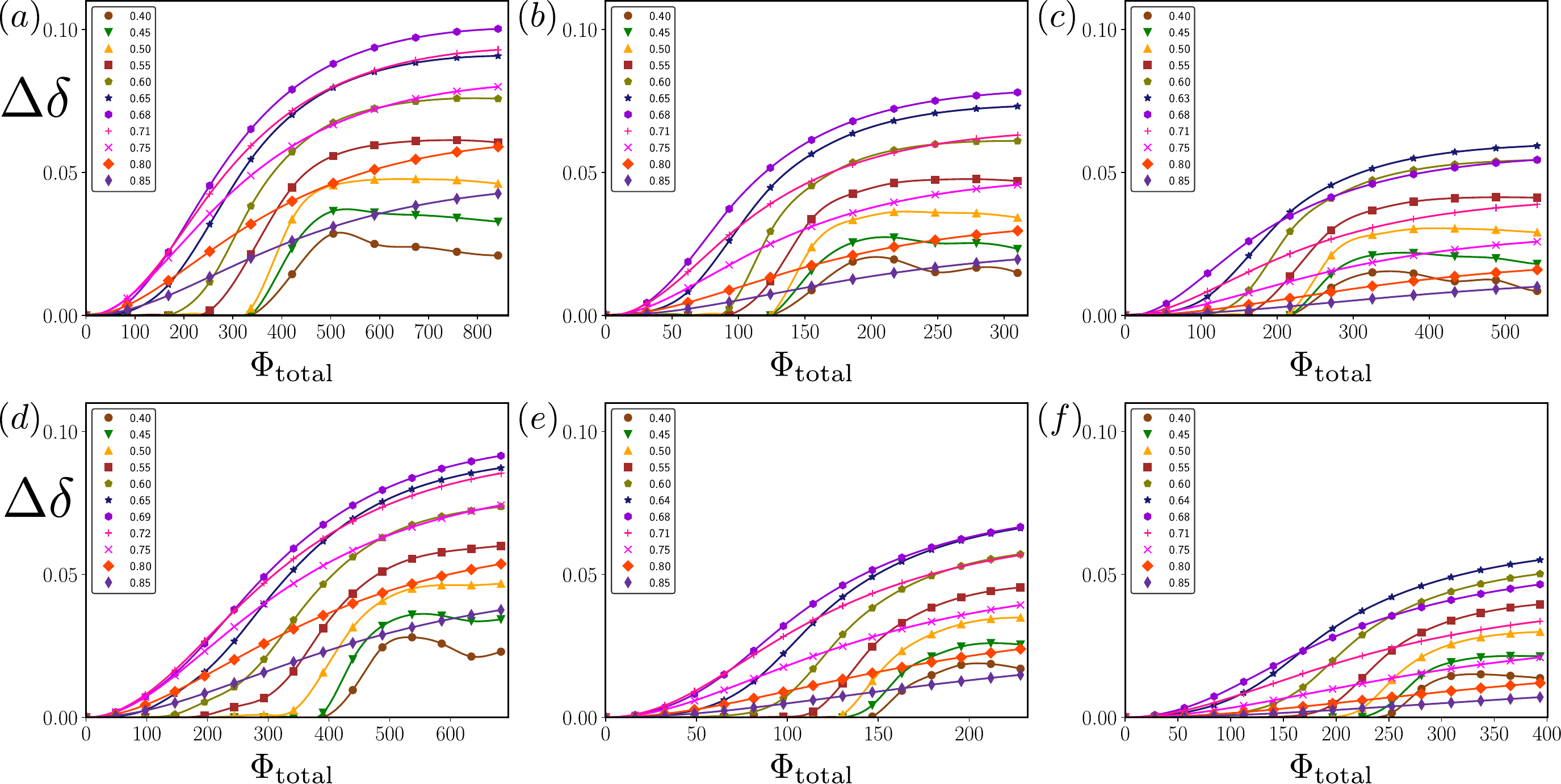}
\caption{The difference between the charge-density-wave orders in the presence [$\delta(\Phi_{\rm total})$] and absence [$\delta(0)$] of magnetic field, defined as $\Delta \delta=\delta(\Phi_{\rm total}) - \delta(0)$ as a function of the total flux enclosed by the system ($\Phi_{\rm total}$) for a wide range (both subcritical and above critical) of the nearest-neighbor Coulomb repulsion ($V$). Positive definite $\Delta \delta$  confirms the curved space magnetic catalysis mechanism in hyperbolic Dirac materials. The results are shown for (a) and (d) (10,3), (b) and (e) (14,3), and (c) and (f) (18,3) hyperbolic lattices in the presence of (a)-(c) uniform and (d)-(f) bell-shaped inhomogeneous magnetic fields. For details see Appendix~\ref{SMsec:additionalnumerics} and Sec.~\ref{subsec:fluxprofile}. 
}~\label{fig:cell}
\end{figure*}
%%%%%%%%%%%%%%%%%%%%%%%%%%%%%%%%%%%%%%%%%%%%%%%%%%%%%%%%%
%%%%%%%%%%%%%%%%%%%%%%%%%%%%%%%%%%%%%%%%%%%%%%%%%%%%%%%%%
%%%%%%%%%%%%%%%%%%%%%%%%%%%%%%%%%%%%%%%%%%%%%%%%%%%%%%%%%
%%%%%%%%%%%%%%%%%%%%%%%%%%%%%%%%%%%%%%%%%%%%%%%%%%%%%%%%%
%%%%%%%%%%%%%%%%%%%%%%%%%%%%%%%%%%%%%%%%%%%%%%%%%%%%%%%%%

Designer electronic materials~\cite{ManoharanNat2012, ManoharanReview2013, GomesNatCom2017, MoraisSmithNatPh2019} and cold atomic setups~\cite{EsslingerReview2010, ZwierleinScience2016, MarkusGreinerNature2017} are two promising platforms where our predicted magnetic catalysis of CDW order in HDMs can be experimentally observed. A hyperbolic designer material can be created by growing its substrate on a suitable material with a different thermal expansion coefficient, such that under cooling a curved substrate is generated. It then can be decorated with the sites of the desired HDM. When placed in strong magnetic fields, designer HDMs can exhibit magnetic catalysis of dynamic mass generation. In cold atomic setups, hyperbolic tessellations can be achieved by suitable arrangements of the laser traps and magnetic fields can be introduced through the coupling of neutral fermions with synthetic gauge fields~\cite{opticalLL1, opticalLL2} to showcase the magnetic catalysis of CDW. As the magnetic catalysis of the CDW order by external magnetic fields occurs in the entire hyperbolic lattice (bulk and boundary) with the open boundary condition, whose magnitude increases with increasing field strength and interaction (see Figs.~\ref{fig:Scaling} and~\ref{fig:gendependence} and the discussion in Appendix~\ref{SMsec:additionalnumerics}), the predicted outcomes should be observable in real materials, which are always synthesized with open boundaries.

\acknowledgments

B.R.\ was supported by NSF CAREER Grant No.\ DMR-2238679. B.R.\ thanks Ariel Sommer, Abhisek Samanta, and Sourav Manna for discussions, and Vladimir Juri\v{c}i\'c for a critical reading of the manuscript. Contributions of Noble Gluscevich in the early stage and Christopher Leong in the final stage of this project are gratefully acknowledged.

\appendix

\section{Details of DOS Calculation}~~\label{SMsec:DOSdetails}

In this appendix, we present the details of the DOS calculations using Gaussian smoothing. The formula for the DOS is then given by 
\begin{equation}
\rho(E) = \frac{1}{N} \sum_j \frac{1}{\sqrt{2 \pi \sigma^2}}\exp\left[ -\frac{(E-\epsilon_j)^2}{2 \sigma^2} \right],
\end{equation}
where $N$ is the total number of sites on the hyperbolic lattice, $\epsilon_j$ is the energy eigenvalue, and $\sigma$ is the smoothing parameter. For Fig.~\ref{fig:DOS}, $\sigma$ is equal to 0.08 for the $(10,3)$ hyperbolic lattice and 0.06 for the $(14,3)$ and $(18,3)$ hyperbolic lattices. These values are large, so it is necessary to subtract the background DOS at zero energy in the absence of magnetic fields. For the DOS computed in Fig.~\ref{fig:DOS_gap}, the smoothing parameter $\sigma$ is equal to 0.015 for the $(10,3)$ and $(14,3)$ hyperbolic lattices, and $\sigma = 0.0105$ for the $(18,3)$ hyperbolic lattice.

In Fig.~\ref{fig:DOS}, the quadratic dependence of the DOS near zero energy with the total flux enclosed by the system is reported using Gaussian smoothing. However, even without such smoothing the same scaling behavior is observed, as shown in Fig.~\ref{fig:DOSappend}.

\section{Additional Numerical Results}~\label{SMsec:additionalnumerics}

In this appendix, we present some additional numerical results to unambiguously establish the curved space magnetic catalysis mechanism of HDMs.

To show that for a subcritical strength of the nearest-neighbor Coulomb repulsion ($V<V_c$) the CDW order develops in the \emph{entire} system when it is subject to an external magnetic field, we define a \emph{local} CDW order-parameter over the sites belonging to the $n$th generation of the hyperbolic lattice containing $N_n$ sites, according to [similar to Eq.~\eqref{eq:CDWOP}] 
\begin{equation}~\label{eq:CDWOPLocal}
\delta_n=\frac{1}{N_n} \bigg( \sum_{j \in n} \delta_{A,j} + \sum_{j \in n} \delta_{B,j} \bigg). 
\end{equation}
The variations of $\delta_n$ with $n$ when the (10,3), (14,3) and (18,3) hyperbolic lattices are immersed in uniform and inhomogeneous magnetic fields for subcritical strengths of the nearest-neighbor Coulomb repulsion are shown in Fig.~\ref{fig:gendependence}. Figure~\ref{fig:gendependence} shows that $\delta_n$ is finite for any $n$, and for different values of $n$, all $\delta_n$ are comparable, establishing that the magnetic catalysis mechanism is operative over the entire system.

A discussion of the spatial variation of the CDW order in the presence of external magnetic fields in HDMs with open boundary conditions is due at this point. Notice that in the presence of uniform magnetic fields, the CDW remains \emph{almost} constant in the bulk of the system, while it dips slightly at the edges, but remains finite and comparable to that in the bulk, as shown in Figs.~\ref{fig:gendependence}(a)-\ref{fig:gendependence}(c). On the other hand, when HDMs are subject to bell-shaped inhomogenous magnetic fields, whose strength decreases almost linearly from the bulk to the edge of the system (see Sec.~\ref{subsec:fluxprofile} for details), the CDW order parameter also decreases almost linearly, as shown in Figs.~\ref{fig:gendependence}(d)-\ref{fig:gendependence}(f). In this case, the local CDW order closely follows the profile of the external magnetic field, while remaining finite everywhere in the system. In addition, the amplitude of the CDW order increases monotonically with increasing strength of the subcritical NN repulsion $V$, making them observable in real systems and the thermodynamic limit with open boundary conditions. These outcomes are qualitatively similar to the ones, previously reported for honeycomb lattice, hosting massless Dirac fermions on flat space, subject to external magnetic fields with open boundary condition (see Fig.~8 in Ref.~\cite{RoyHerbutPhysRevB2011}). Given that the scaling of the interaction-induced gap, computed within the Hartree approximation, following the magnetic catalysis mechanism successfully explained the experimentally observed scaling of interaction-induced gap within the zeroth Landau level of graphene~\cite{RoyKennettDassarma2014}, this mechanism should also be operative in HDMs and the weak-coupling instability of the CDW order in the presence of external magnetic field should be observable in real materials in the thermodynamic limit.

Finally, note that numerical estimation of the zero-magnetic-field critical strength of the nearest-neighbor Coulomb repulsion ($V_c$) for the CDW ordering from a real space Hartree-Fock analysis is affected by the finite-size effect. We therefore compute the quantity $\Delta \delta$, the difference between the magnetic-field induced CDW order ($\delta(\Phi_{\rm total})$) and the zero-field CDW order ($\delta(0)$) as a function of the total flux enclosed by the system $\Phi_{\rm total}$ for a wide range of nearest-neighbor Coulomb repulsion, including both $V< V_c$ and $V>V_c$. For any choice of $V$, we find that $\Delta \delta$ is a positive definite quantity, as shown in Fig.~\ref{fig:cell}, endorsing the curved space magnetic catalysis.

%%%%%%%%%%%%%%%%%%%%%%%%%%%%%%%%%%%%%%%%%%%%%%%%%%%%%%%%%%%%%%%%%%%%%%
\bibliography{HyperbolicReferences} 
%%%%%%%%%%%%%%%%%%%%%%%%%%%%%%%%%%%%%%%%%%%%%%%%%%%%%%%%%%%%%%%%%%%%%%

\end{document}